\documentclass[9pt,twocolumn,twoside,lineno]{pnas-new}

\templatetype{pnasresearcharticle} 

\title{Intersectional Synergies: Untangling Irreducible Effects of Intersecting Identities via Information Decomposition}

\author[1,2]{Thomas F. Varley}
\author[1,3]{Patrick Kaminski} 

\affil[1]{School of Informatics, Computing, and Engineering, Indiana University, Bloomington, IN, USA}
\affil[2]{Department of Psychology \& Brain Sciences, Indiana University, Bloomington, IN, USA}
\affil[3]{Department of Sociology, Indiana University, Bloomington, IN, USA}

\leadauthor{Varley} 

\significancestatement{\textit{Intersectionality} has become a frequent topic of discussion both in academic sociologyand among popular movements for social justice such as Black Lives Matter and LGBT rights. Intersectionality proposes that an individual's experience of society is irreducible to the sum of one's various identities considered individually, but are ``greater than the sum of their parts." By using techniques from information theory (partial information decomposition) we can identify intersectional relationships in empirical data in a manner not possible with standard multiple-linear regression. We show that robust synergistic intersectional relationships exist between identities such as how Race and Sex influence income, and that these synergies are stable over time. This work opens new avenues for researching higher-order relationships in social processes.}

\authorcontributions{TFV conceptualized the project, performed data analysis, and drafted the manuscript. PK assisted with interpretation and editing the manuscript.}
\authordeclaration{The authors declare no competing interests.}

\correspondingauthor{\textsuperscript{2}To whom correspondence should be addressed. E-mail: tvarley@iu.edu}

\keywords{Intersectionality $|$ Information Theory $|$ Social Justice $|$ Complex Systems $|$ Partial Information Decomposition} 

\begin{abstract}
\textit{Intersectionality} has become a frequent topic of discussion both in academic sociologyand among popular movements for social justice. Intersectionality proposes that an individual's experience of society has aspects that are irreducible to the sum of one's various identities considered individually, but are ``greater than the sum of their parts." In this work, we show that the effects of intersectional identities can be statistically observed in empirical data using information theory. We show that, when considering the predictive relationship between various identities categories such as race, sex, and income on outcomes such as health and wellness, robust statistical synergies appear. These synergies show that there are joint-effects of identities on outcomes that are irreducible to any identity considered individually and only appear when specific categories are considered together (for example, there is a large, synergistic effect of race and sex considered jointly on income irreducible to either race or sex). We then show using synthetic data that the current gold-standard method of assessing intersectionalities in data (linear regression with multiplicative interaction coefficients) fails to disambiguate between truly synergistic, greater-than-the-sum-of-their-parts interactions, and redundant interactions. We explore the significance of these two distinct types of interactions in the context of making inferences about intersectional relationships in data and the importance of being able to reliably differentiate the two. Finally, we conclude that information theory, as a model-free framework sensitive to nonlinearities and synergies in data, is a natural method by which to explore the space of higher-order social dynamics. 
\end{abstract}

\dates{This manuscript was compiled on \today}
\doi{\url{www.pnas.org/cgi/doi/10.1073/pnas.XXXXXXXXXX}}

\begin{document}

\maketitle
\thispagestyle{firststyle}
\ifthenelse{\boolean{shortarticle}}{\ifthenelse{\boolean{singlecolumn}}{\abscontentformatted}{\abscontent}}{}

"Intersectionality" refers to the the idea that an individual's experience of social privilege or oppression is a function of how all of the various identities (such as race, sex, class, ability, etc.) that a given person can hold \textit{intersect,} and that the result of these intersecting identities is not directly decomposable into the sum of all identities considered individually. Intersectionality as a framework was first articulated by Kimberle Crenshaw \cite{crenshaw_demarginalizing_1989,crenshaw_mapping_1991} who discussed how Black women in particular face distinct forms of marginalization in the context of the justice system. She argued that pre-existing theoretical frameworks of both feminist and anti-racist scholarship erased the specific experiences of Black women, instead operating under a framework where "All the Women Are White; All the Blacks Are Men" \cite{crenshaw_demarginalizing_1989}. The particular intersection of Blackness and race has been called \textit{misogynoir} by Moya Bailey \cite{bailey_misogynoir_2021} to articulate its status as a unique, irreducible social experience. 

From its initial focus on Black feminism and the unique experience of Black women, the field of Intersectionality Studies has expanded \cite{cho_toward_2013} and now many different axes of identity are routinely studied, including sexuality \cite{taylor_theorizing_2010,ferlatte_recent_2018}, class \cite{block_exploring_2014}, disability \cite{nirmala_unspeakable_2010,ben-moshe_introduction_2014}, and/or immigration status and citizenship \cite{erez_intersectionality_2018,romero_inclusion_2008,viruell-fuentes_more_2012,yuvaldavis_intersectionality_2007}. Outside of academia, intersectionality has gained prominence in popular culture, particularly around issues related to the Black Lives Matter movement, trans and gender-nonconforming rights, and the issue of immigration from South and Central America into the United States. As a consequence of the growing public discussion and activism, there has been a growing interest in how intersectional frameworks could be deliberately incorporated into public policy decisions \cite{hankivsky_intersectionality_2011,hankivsky_introduction_2019,garcia_incorporating_2021}. As theories of intersectionality leave the Academy and begin to influence more mainstream decision-making, it is imperative that the field develops analytic tools that allow for the discussion of intersecting identities in the context of empirical data. Understanding the causal and material impacts of policy in the context of intersectional identities requires the development of robust empirical methodologies that can be used to inform and identify interventions. 

Previous researchers have highlighted the difficulties in addressing the issue of synergistic relationships between identities in empirical data: for example Bowleg concisely summarizes the issue as ``Black + Lesbian + Woman $\not=$ Black Lesbian Woman", and discusses the problems with assuming additive relationships, although Boweleg is primarily concerned with analyzing qualitative ethnographic data and does not engage with the mathematical issues around super-additive relationships in numerical or categorical data \cite{bowleg_when_2008}. There has been discussion of how intersectionality can be accounted for by quantitative methods, for example Scott and Siltanen \cite{scott_intersectionality_2017} discuss multi-level linear regressions informed by context, and Rouhani provides a gold-standard primer distinguishing between additive and multiplicative effects using linear regressions \cite{rouhani_intersectionality-informed_2014}. The standard practice is associating intersectional effects with the multiplicative interaction term in linear regression and comparing it to the main effect using estimators of prediction error like Akaike's Information Criteria. 

While powerful, these methods have a number of subtle limitations that can complicate the analysis of complex data. The first, and most glaring, is that the reliance on parametric models, goodness-of-fit tests, and arbitrary criteria such as the $\alpha$-level of statistical significance can limit the kinds of relationships the analysis is sensitive to. For example, the reviewed literature makes overwhelming use of \textit{linear} (Gaussian) assumptions in assessing main and interaction effects and does not account for non-linear relationships between interacting variables. A more subtle concern is that, while linear regressions can compare main and interaction effects, and in doing so account for the fact that identities can interact, it fails to capture the synergistic core of intersectionality. For example, a Black woman plausibly experiences: generic anti-Black racism (independent of sex), generic misogyny (independent of race), and intersectional misogynoire specific to her identities as a Black woman (there may also be ``redundant" effects: a cost shared by Blacks and women by virtue of being a minority of any type and not specific to either identity). Linear regression models fail to decompose this constellation of effects as it has no way to rigorously account for, or even acknowledge, differences between redundant and synergistic relationships. To address these issues, we turn to an alternative statistical framework for analyzing data: information theory.

Information theory represents an appealing statistical framework with which to tackle this issue: in contrast to standard linear regression models, information theory is almost entirely model-free, making it sensitive to non-linear relationships in data \cite{cover_elements_2012}. More generally, being based purely on joint and conditional probability distributions, information theory is ``epistemically modest", and deeply linked to the general process of making inferences under conditions of uncertainty, a key concern when assessing complex systems \cite{mackay_information_2003}. Finally, for researchers who do wish to leverage the power of linear models, or work with continuous data, closed-form Gaussian estimators of all major information-theoretic relationships exist \cite{cover_elements_2012}, as do non-parametric, continuous estimators based on K-nearest neighbor relationships \cite{kraskov_estimating_2004} for non-linear analysis of real-valued (continuous) data (also see \cite{lizier_jidt_2014}, Supplementary Material). 

Crucially, information theory is well-equipped to handle the problem of decomposing multivariate relationships in data into synergistic (intersectional) and redundant components using a framework known as partial information decomposition (PID) \cite{williams_nonnegative_2010,gutknecht_bits_2020} (see Section \ref{sec:pid} for details), and has been applied in a variety of fields, including interpretable machine learning \cite{tax_partial_2017}, medical imaging \cite{colenbier_disambiguating_2020}, biological neural networks \cite{timme_high-degree_2016,faber_computation_2018}, ecology \cite{goodwell_temporal_2017}, evolution \cite{luppi_synergistic_2020}, as well as to philosophical questions such as the nature of ``emergence" \cite{rosas_reconciling_2020,varley_causal_2020} and consciousness \cite{luppi_synergistic_2020-1}. This interdisciplinary group of results suggests that synergistic relationships ``greater than the sum of their parts" are ubiquitous in both natural and human-made systems, so it is natural to hypothesize that they may also exist in social systems. 

We hypothesize that intersectional inequalities (i.e. costs associated with being both Black and female or advantages associated with being both White and male) should be observable in population-level demographic and life outcomes data using information theoretic analysis, and that information theory will out-perform linear regression in discriminating between redundant and synergistic effects. 

\section{Methods}
\subsection{Basic Information Theory}
Information theory is a mathematical framework that describes how different interacting entities inform on and constrain each-other's behavior \cite{cover_elements_2012}. Originally developed in the context of theories of communications \cite{shannon_mathematical_1948}, information theory has become an indispensable tool for the analysis of complex systems \cite{lizier_local_2013,bossomaier_introduction_2016}. The core object of study in information theory is the \textit{entropy}, which quantifies how uncertain we, as observers, are about the state of a variable we are observing. Inference, then, is typically understood as the process of minimizing entropy by understanding how information about the variable in question is disclosed by both it's own statistics, and related variables. Given a variable $X$ which can take on different states \textit{x} drawn from a support set $\mathcal{X}$ with probability distribution $P_X(x)$, the entropy of the variable is given by:

\begin{equation}
    H(X) = -\sum_{x\in\mathcal{X}}P_X(x)\log(P_X(x))
\end{equation}

If we are observing more than one variable, the joint entropy is an easy generalization: for a pair of variables $\{X,Y\}$, the joint entropy is given by:

\begin{equation}
    H(X,Y)=-\sum_{x\in\mathcal{X}}\sum_{y\in\mathcal{Y}}P_{X,Y}(x,y)\log(P_{X,Y}(x,y))
\end{equation}

Which quantities the total uncertainty about the state of both variables simultaneously. From the joint and marginal entropies we can calculate the conditional entropy:

\begin{equation}
    H(X|Y) = H(X,Y) - H(Y)
\end{equation}

This conditional entropy tells us how much uncertainty about the state of $X$ remains after we have ``accounted for" the state of $Y$. It is important to note that $H(X|Y) \leq H(X)$: information about the state of $Y$ can only ever reduce our uncertainty about the state of $X$ or, if $X$ and $Y$ are independent, provide no insight. 

So far, all the measures that we have discussed have been measures of uncertainty, \textit{not} measures of information. To quantify information itself, we introduce the mutual information:

\begin{equation}
    I(X;Y) = H(X) - H(X|Y)
\end{equation}

\begin{figure}
    \centering
    \includegraphics[scale=0.65]{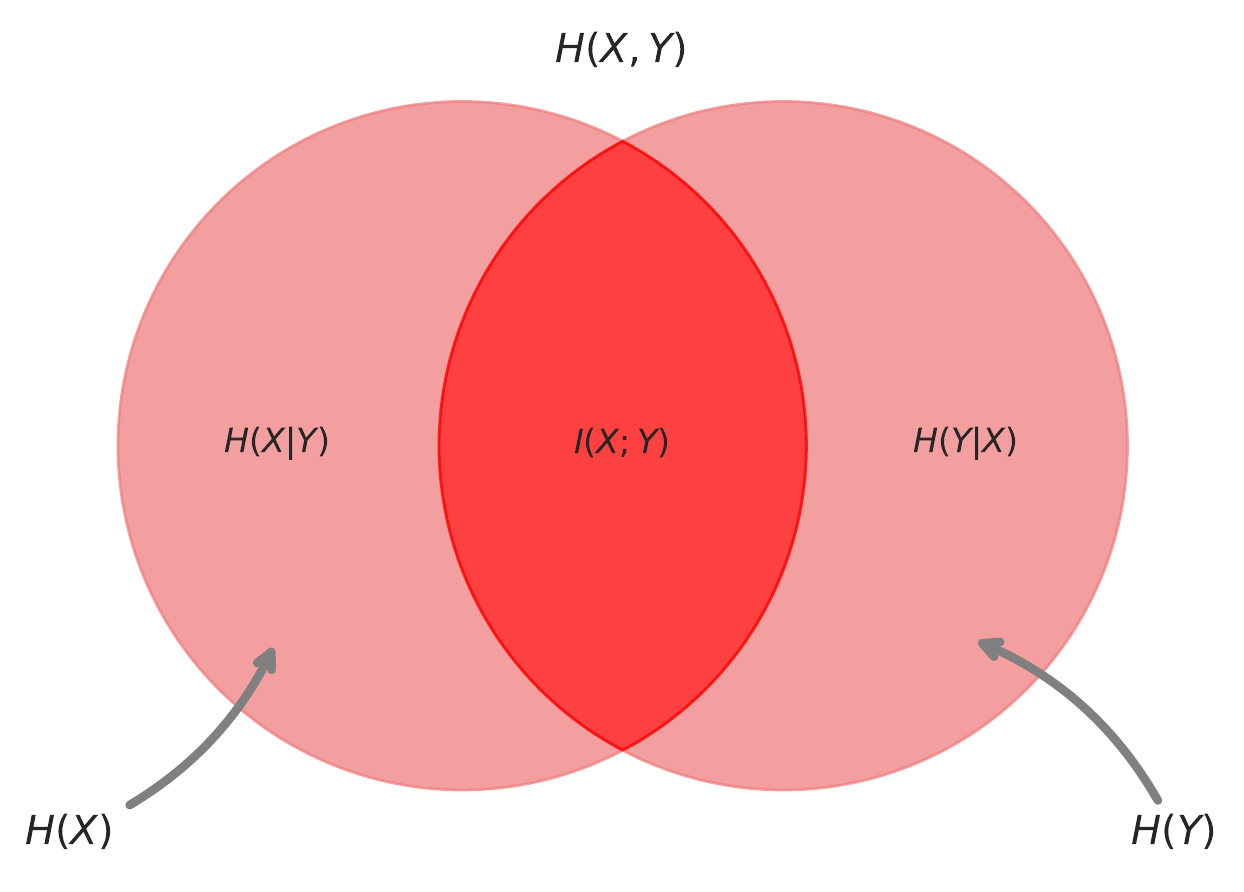}
	\caption{\textbf{Mutual information as the intersection of marginal entropies.} Mutual information can be understood as the intersection the entropies of two correlated variables. This plot highlights the intuition that $I(X;Y)=H(X)-H(X|Y)=H(Y)-H(Y|X)$. Venn diagrams made using the excellent matplotlib-venn package.}
	\label{fig:mi}
\end{figure}

For a visual intuitive aid, see Figure. \ref{fig:mi}

It is worth unpacking this to build intuition: we begin with our uncertainty about the state of $X$, given by $H(X)$. From this, we subtract our remaining uncertainty about $X$ after the state of $Y$ has been taken into account ($H(X|Y)$). The difference of these two quantities is our uncertainty about $X$ that is \textit{resolved} by knowledge of $Y$: the amount of information $Y$ provides about $X$. It is important to note that the mutual information is symmetric: $I(X;Y) = I(Y;X)$. As with the joint entropy, it is possible to calculate the mutual information between a set of variables and a single target. Consider the case of trying to predict the state of a variable $Y$ based on two predictor variables $X_1$ and $X_2$. We can calculate $I(X_1,X_2 ; Y)$ the sane way as above, only treating the joint states of $X_1$ and $X_2$ as a single macro-variable. 

It is crucial that the joint mutual information is \textbf{not} always equivalent to the sum of the individual marginal mutual informations:

\begin{equation}
    I(X_1,X_2;Y) \not= I(X_1;Y) + I(X_2;Y)
\end{equation}

 The joint mutual information can be either greater, or less than, the sum of the marginal mutual informations depending on how correlated information is distributed across the two predictor variables. If $I(X_1,X_2;Y) < I(X_1;Y) + I(X_2;Y)$ then the total is less than the sum of their parts and consequently $X_1$ and $X_2$ must share some \textit{redundant} information. Consider a Venn diagram: if the area of two overlapping circles is less than the sum of the areas of both circles, then there is redundant area shared between them. Conversely if $I(X_1,X_2;Y) > I(X_1,Y) + I(X_2,Y)$, then the whole is \textit{greater} than the sum of the parts: there is information about $Y$ in the joint state of $X_1,X_2$ that cannot be extracted from either variable considered independently.

Finally, it is entirely possible that a pair of predictor variables contains \textit{both} redundant and synergistic information: while the difference between the joint and the sum of the marginal mutual informations is a decent heuristic to test if there is any synergy at all in the triad, it does not tell us \textit{how much} redundant or synergistic information is present, nor what proportion of the total information is accounted for by redundant or synergistic information. For this we need more advanced mathematical machinery. 

\subsection{Partial Information Decomposition}
\label{sec:pid}

Given a set of predictor variables influencing a shared predicted variable, the partial information decomposition framework \cite{williams_nonnegative_2010,gutknecht_bits_2020} provides the tools necessary to decompose the joint mutual information into redundant, synergistic, and unique ``types." The resulting decomposition is given by:

\begin{eqnarray}
    I(X_1,X_2;Y) &=& Red(X_1,X_2;Y) + \\ 
    &&Unq(X_1;Y/X_2) + \notag \\ 
    &&Unq(X_2;Y/X_1) + \notag \\ 
    &&Syn(X_1,X_2;Y) \notag
\end{eqnarray}

where $Red(X_1,X_2,Y)$ quantifies the information about $Y$ that can be resolved by observing $X_1$ or $X_2$, $Unq(X_1;Y/X_2)$ quantifies the amount of information about $Y$ is uniquely disclosed by $X_1$ (in the context of $X_2$) and vice versa. Finally, $Syn(X_1,X_2;Y)$ quantifies the information disclosed only by the joint state of $X_1$ and $X_2$, and no simpler combination of variables. 

Furthermore, we can decompose the marginal mutual informations in the same fashion:

\begin{eqnarray}
I(X_1;Y) &=& Red(X_1,X_2;Y) + Unq(X_1;Y/X_2)\\
I(X_2;Y) &=& Red(X_1,X_2;Y) + Unq(X_2;Y/X_1)
\end{eqnarray}

The result is a system if three equations, three known quantities (the joint and marginal mutual informations), and four unknown variables (\textit{Red, Unq$_1$, Unq$_2$, Syn}). If it is possible to identify any one partial information atom, then the remainder are trivial (``three for free"). 

\subsubsection{Choosing a Partial-Information Function}
Unfortunately, classical Shannon information theory does not provide a function that calculates any of these, and neither does the PID framework itself. Much work has gone into developing such functions, and as yet, there is no consensus within the field of information theory as to the ``gold standard." To ensure the robustness of the concept, we analyzed the data using two different functions to ensure that the distributions of redundancy and synergy remained consistent despite different technical starting points. The first was the original measure proposed by Williams and Beer \cite{williams_nonnegative_2010} called $I_{min}$, which gives the redundant information shared by a set of sources about a target as:

\begin{equation}
    I_{min}(X_1, X_2, ...X_n ; Y) = \sum_{y \in \mathcal{Y}}P(Y=y)\min_{i}I(X_i ; Y=y)
\end{equation}

This measure (also called the ``specific information") gives the minimum amount of information that the set of sources discloses about the target. To ensure that our results were robust to the particular free parameters inherent in the PID framework, we replicated our results using the measure of $I_{BROJA}$ proposed by Bertschinger et al., \cite{bertschinger_quantifying_2014,griffith_quantifying_2014}. The $I_{BROJA}$ method starts with the unique information, rather than the redundant information; defining it as the minimum conditional conditional mutual information possible while holding the marginals constant. The extract the unique information $Unq(X_1 ; Y / X_1$, Bertschiner et al., begins by defining the set of all joint distributions of $X_1$, $X_2$, and $Y$ such that the marginals are equivalent to the empirical distribution $P$:

\begin{equation}
    \Delta = \{Q | p(x_i,y) = q(x_i,y) \forall i\}
\end{equation}

The unique $I_{BROJA}$ for a given source is defined as:

\begin{equation}
    I_{BROJA}(X_1 ; Y / X_2) = \min_{Q \in \Delta}I^{Q}(X_1 ; Y | X_2)
\end{equation}

Where $I^{Q}(\cdot ; \cdot | \cdot)$ indicates that the mutual information being calculated with respect to the probability distribution $Q(X_1, X_2, Y)$. Once the two unique informations have been calculated, the rest of the 2-source lattice can be solved with basic algebra. Both $I_{min}$ and $I_{BROJA}$ provide strictly positive values for all partial information atoms on the redundancy lattice, allowing for a complete decomposition of the joint mutual information terms. 

\begin{figure}
    \centering
    \includegraphics[scale=0.55]{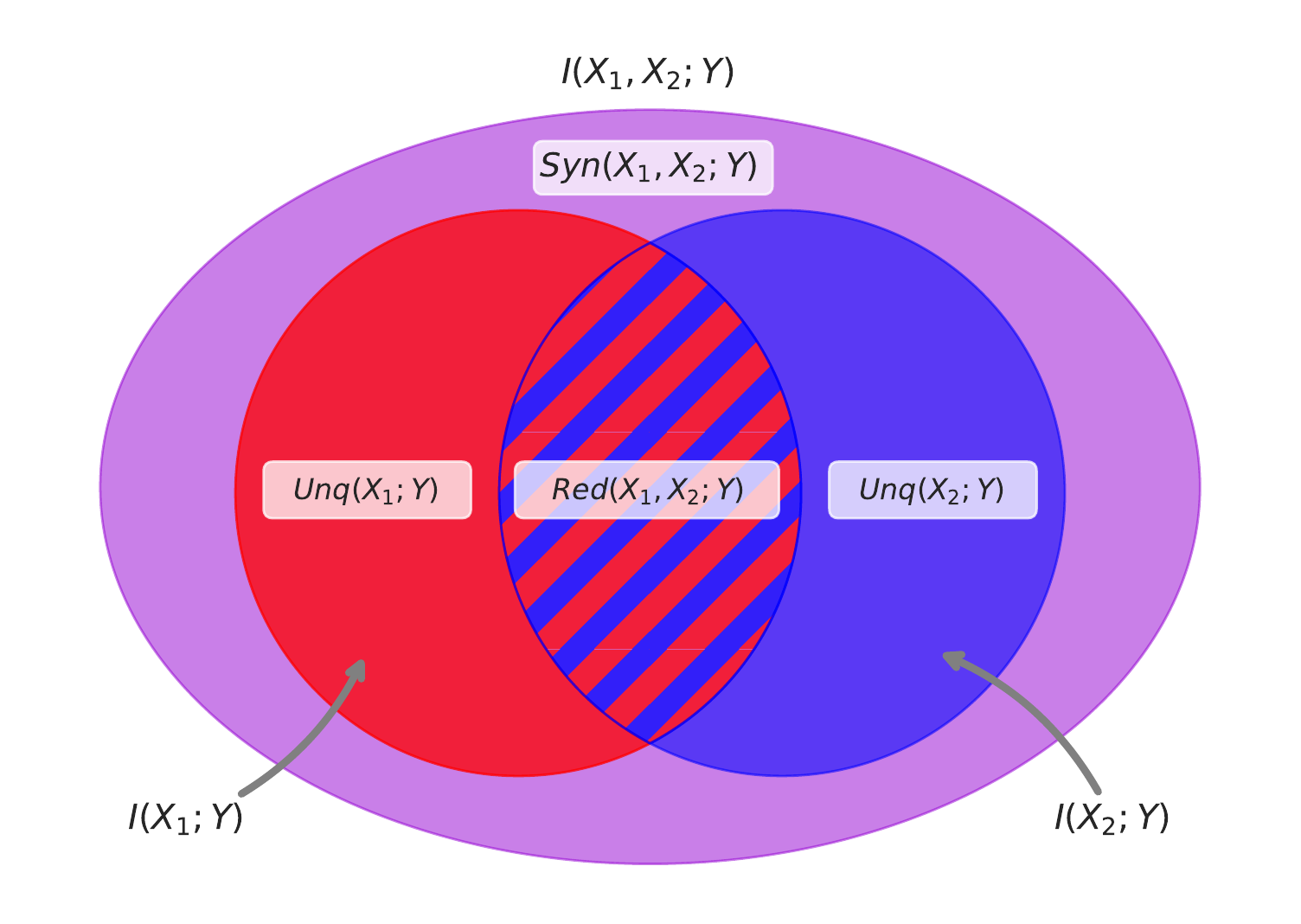}
	\caption{\textbf{Partial information decomposition for two predictor variables and a single predicted variable.} A Venn Diagram showing how the various components of partial information (redundant, unique, and synergistic) are related to the joint and marginal mutual information terms for two source variables $X_1$ and $X_2$, and a target variable $Y$. The two circles correspond to the mutual information between each source and the target, while the large ellipse gives the joint mutual information between both sources and the target.
	Notice that the marginal mutual informations overlap, each one counting the redundant (shared) information towards it's own marginal mutual information. We can also see that $I(X_1,X_2;Y) > I(X_1;Y)\cup I(X_2;Y)$: the difference is the synergistic information which cannot be resolved to either marginal mutual information.}
	\label{fig:pid_venn}
\end{figure}

\subsection{Data \& Pre-processing}
For this project, we used a decade of data (2011 - 2020) from the Annual Social and Economic (ASEC) Supplement (available \href{https://www.census.gov/data/datasets/time-series/demo/cps/cps-asec.2020.html}{here}), which provides individual-level micro-data on demographic and economic indicators in the United States. We excluded individuals younger than 25 years of age, or older than 65 years of age as they were less likely to be working full-time (children, students, or retired), individuals not native-born US Citizens, and we only analyzed those who self-identified as "White only" or "Black only" to limit the confounds associated with multi-racial identities. Finally, we only considered individuals who were working in full-time employment since we were looking explicitly at the effects of identities like race or sex on income and the possibility of different levels of employment would represent a possible confound. . 

To limit the size of the joint probability space, we coarse-grained the income distribution: individuals making \$0.00 - \$27,499 were placed into a Low Income group, individuals making  \$27,500 - \$52,499 where placed into a Lower-Middle Income group, individuals making \$52,500 - \$77,499 where placed into Upper Middle Income group, and finally those making more than \$77,500 were placed into an High Income group, although we found that our results were generally robust to the exact choice of cutoffs or the number of categories up to a point at which point the joint probability distribution was too sparsely sampled.

\subsection{Analysis Pipeline}
Performing the PID analysis requires constructing the entire joint-probability distribution for all variables. For example, when untangling the synergistic effects of sex and race on income, it is necessary to compute:

\begin{centering}
\begin{eqnarray}
&P(Race=Black \wedge Sex=Male \wedge Income=Low) \nonumber \\ 
&P(Race=White \wedge Sex=Male \wedge Income=Low) \nonumber \\
&P(Race=Black \wedge Sex=Female \wedge Income=Low) \nonumber \\
&... \nonumber \\
&P(Race=white \wedge Sex = Female \wedge Income=High) \nonumber
\end{eqnarray}
\end{centering}

where $\wedge$ is the logical-AND operator. 

To accurately estimate these probabilities, it is crucial to ensure that the distribution of the joint states of the predictor variables (in this case, race and sex) are evenly distributed: there should be an equal number of Black women, White women, Black men, and White men. If, for example, White men are over-represented and have higher-then-average incomes, which is the case in the full dataset, then that will incorrectly skew the influence of White male data points on the joint distribution and expected moments. To address this issue, we created 1000 sub-sampled distributions while keeping the number of the joint-states of the predictive variables the same (i.e. the same number of Black women, White women, Black men, and White men). We then calculated the mixture distribution of all 1000 sub-sampled distributions and used that to perform the analysis. By resampling the distribution to enforce a maximum-entropy distribution, we are controlling for the problem of uneven sampling, such as oversampling done by the census-takers to ensure sufficient coverage.

By forcing a ``flat" probability distribution on the space of joint identities, this can be understood as being a ``causal" analysis in the sense of Pearl's \textit{do}-calculus model of causality \cite{pearl_causal_2010,pearl_causal_2016} and Woodward's interventionist causal framework \cite{woodward_making_2005}. Specifically, we are are leveraging the measure of ``effective information" (first introduced by Tononi and Sporns \cite{tononi_measuring_2003}) which quantifies all of the causally-relevant relationships between two variables:

\begin{equation}
    EI(X \to Y) = I(X^{H_{max}} ; Y)
\end{equation}

where $X^{H_{max}}$ indicates that the probability distribution of the  of states of the predictor variable $X$ are maximally entropic (in this case, uniform). The maximum entropy distribution is considered to be ``causal" because it controls for any biases that might be introduced from the empirical distribution and preserves only the core ``effective" relationships in the data. The consequence is that, by strictly forcing our distribution of identities to be uniform (i.e. P(Race = Black $\wedge$ Sex = Female) = P(Race = White $\wedge$ Sex = Male)), we are decomposing the effective information (i.e. the ``causally relevant" information) shared between identity and outcomes (as recorded in this data). This is distinctly different from standard, regression-approaches, which typically do not force causal distributions and are consequently only correlational in nature. 

We compared three different sets of variables: \textit{I(Race, Sex ; Income), I(Sex, Income ; Health) and I(Race, Sex, Income ; Health)}, which represent a sampling of the interactions of three major identity groups (with income standing in as a proxy for class). We began by comparing the difference between the sum of the marginal mutual informations to the joint mutual information as a heuristic to determine whether the relationship was generally synergy dominated (if the sum of the marginals is less than the joint) or redundancy dominated (if the sum of the marginals is greater than the joint). Subsequently we did the full partial information decomposition to extract the relevant partial information atoms. To compare between datasets, each partial information atom was normalized by the joint mutual information to give a proportion of the total information (i.e. of all the information that race and sex provide about class, what proportion of that information is synergistic, redundant, unique, etc). 

Finally, we show that redundant and synergistic information make distinctly different predictions about the effects of identity that can be understood in the context of logical conjunction and disjunction operators, and discuss how these distinct modes of interaction can form a basis for cross-identity solidarity and exclusive intersectional experiences respectively.  

\subsection{Data \& Code Availability}
All data can be downloaded from the \href{https://www.census.gov/data/datasets/time-series/demo/cps/cps-asec.2020.html}{US Census Bureau website}. All scripts necessary to recreate the analysis and figures are included as Supplementary Material. Data cleaning was done using the Pandas and Numpy packages and all information theoretic analysis was carried out in Python, using the \textit{Discrete Information Theory} toolbox by James et al., \cite{james_dit_2018}. 

\section{Results}

\begin{figure}
    \centering
    \includegraphics[scale=0.6]{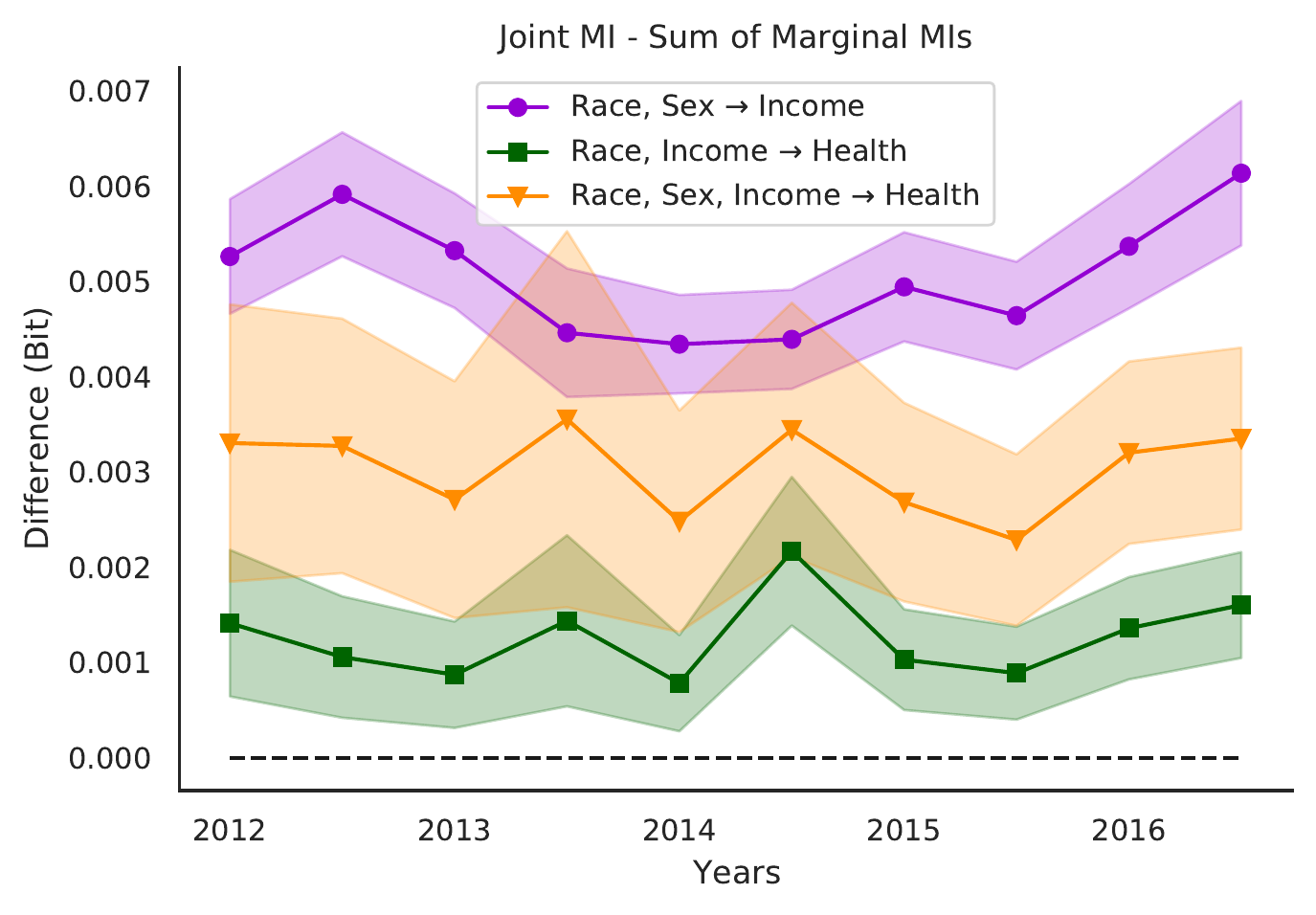}
    \caption{\textbf{The difference between the joint and the sum of the marginal mutual informations for three relationships.} Across all ten years, for all three relationships assessed, the difference between the joint and the sum of the marginal mutual informations was consistently greater than 0, indicating that the joint state of all identity groups considered together disclosed more information about the outcomes (income, health status) then all the identities considered independently. }
    \label{fig:whole-sum}
\end{figure}

We began by examining the difference between the joint mutual information and the sum of the marginal entropies for each of our three relationships of interest across the whole decade (Fig. \ref{fig:whole-sum}). This is a practical heuristic sometimes called the "Whole Minus Sum" method \cite{griffith_quantifying_2014}. We found that, for all three relationships and across all ten years, the joint entropy was greater than the sum of the marginals, indicating that the relationship between various intersecting identities and target outcomes are consistently ``greater than the sum of their parts," or in the context of PID, ``synergy-dominated."  

\begin{figure}
    \centering
    \includegraphics[scale=0.4]{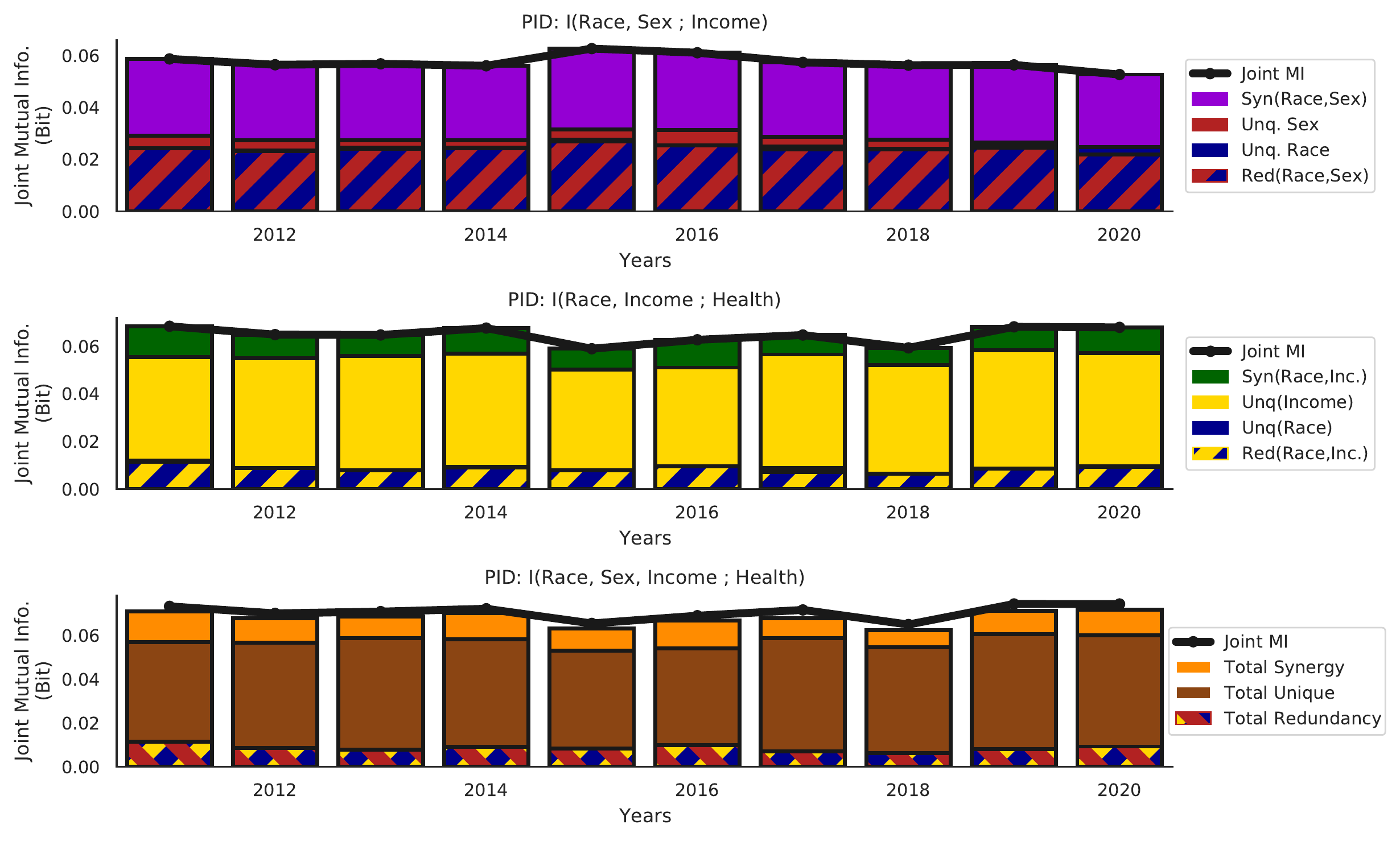}
    \caption{\textbf{Partial information decomposition of I(Race, Sex ; Income), I(Race, Income ; Health Status) and I(Race, Sex, Income ; Health Status). Top:} The PID for race and sex on income. The information about income disclosed by race and sex is almost entirely either redundant or synergistic, race or sex individually disclose very little unique information. \textbf{Middle:} PID for race and income on health status. In contrast to the effects of race and sex, almost all information about health status is disclosed by income, although there is a non-trivial amount of both shared and synergistic information. This shows that interacting identities can have markedly different structures only revealed by information decomposition. \textbf{Bottom:} The PID for the relationship between race, sex, and income on health status. Due to the large number of atoms, we aggregated all purely redundant terms, all purely unique terms, and all purely synergistic terms. The small difference between this total and the complete joint mutual information corresponds to exotic, higher-order interactions not reported here. We can see that, for all three relationships, the degree of informativeness remains remarkably constant over the decade, and that the overall degree of synergy, redundancy, and unique information is similar consistent.}
    \label{fig:pid_bars}
\end{figure}

To determine the exact distributions of redundant, unique, and synergistic information, we then computed the full PIDs (for visualization of the two triadic interactions, see Figure \ref{fig:pid_bars}). We found that the relationship between race and sex on income had a high degree of both synergy and redundancy, with only limited unique information disclosed by only race or sex. On average, $51\pm1.3\%$ of the information disclosed by race and sex about income was synergistic in nature, while $42\pm0.8\%$ of the information was redundantly shared between the two predictor variables. The two unique components together disclosed only $7\pm2\%$ bit total. This contrasts dramatically with the information that race and income provide about general health status: only $15.2\pm2.2\%$ bit of information were synergistic, and only $13.27\pm1.8\%$ was redundantly shared. Income uniquely disclosed $71.13\pm3.88\%$ of the information about health, with only $0.4\pm0.82\%$ being uniquely disclosed by race. Despite the clear differences in the breakdown of partial information, the joint mutual informations were very similar: \textit{I(Race, Sex ; Income)} = 0.057 bit and \textit{I(Race, Income ; Health)} = 0.06 bit - it is only when breaking down the information into it's atomic components that we see meaningful differences. When considering the three-way effect of race, sex, and income on health, we found that $12.1\pm1.9\%$ of the total information was some form of redundant (\textit{Red(Race, Sex, Income), Red(Race, Sex), Red(Race, Income), Red(Race, Sex)}), while another $15.48\pm2.31\%$ was some form of synergistic. Unique information account for $68.88\pm3.76\%$ of the total. Note that these do not sum up to 100\% since there is a small amount of information distributed over more exotic and hard-to-interpret partial information atoms.  

When considering all of the years individually, it is clear that the informational relationships are stable over time (as evidenced by the comparatively low standard deviation terms), There does not appear to be a substantive increase, or decrease in how the various identities interact, suggesting that the intersectional relationships are consistent over multiple years. 

\subsection{Untangling Redundant \& Synergistic Information}

\begin{figure}
    \centering
    \includegraphics[scale=0.35]{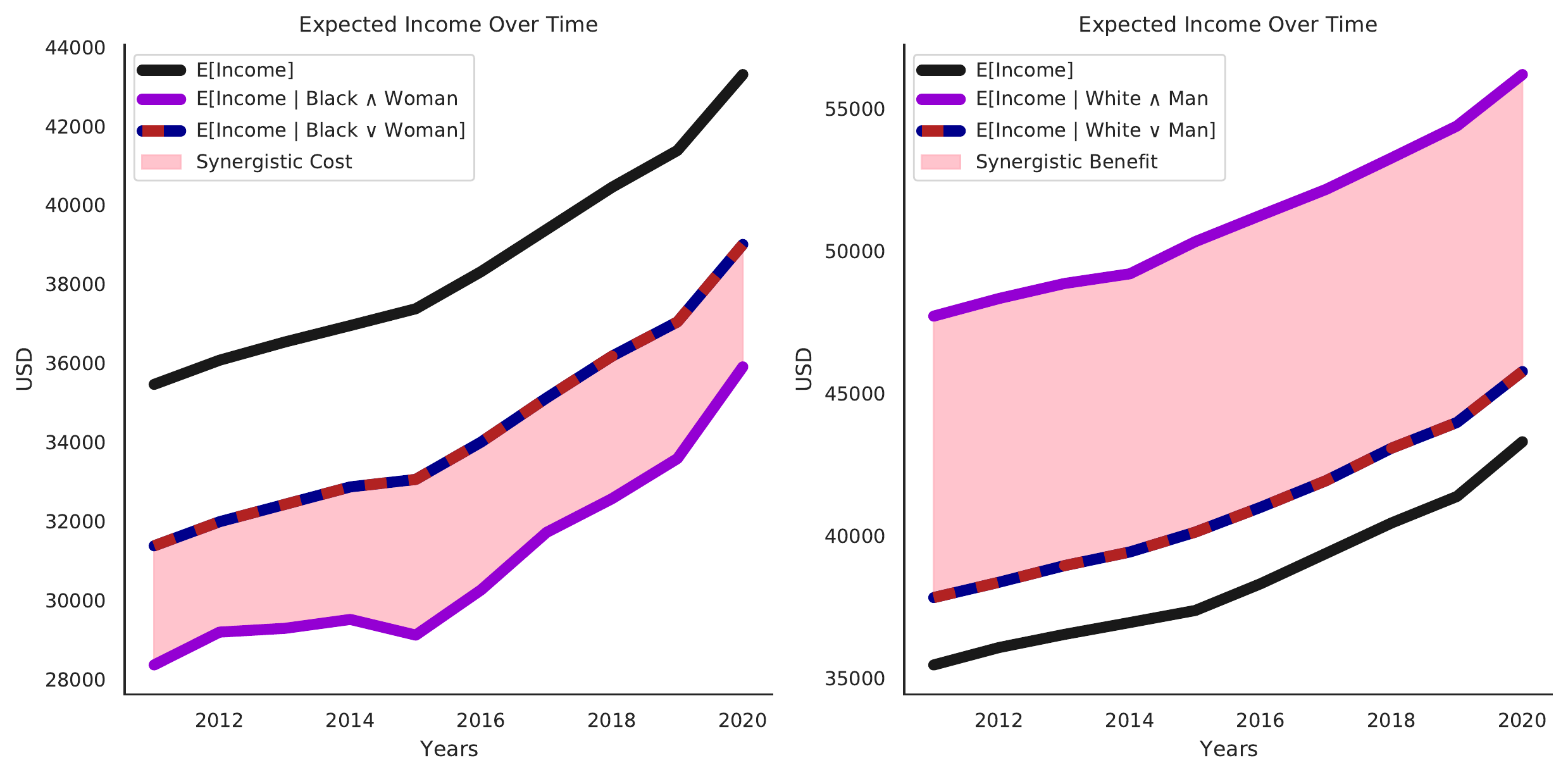}
    \caption{\textbf{The difference between the shared cost of marginalization, and the extra cost of synergistic intersectionality.} \textbf{Left:} The expected income for the whole population over time, the expected income for that subset of the population that is Black or a woman, and, and the expect income for Black women. We can see that there is an extra cost incurred by being Black and a woman that is ``above and beyond" the cost incurred by being Black or a woman. \textbf{Right:} the same plot, but considering Whiteness and masculinity. Once again, the joint relationship is strong different from the disjunction, only this time there is a synergistic benefit to being White and male, as opposed to a synergistic cost.}
    \label{fig:cost-benefit}
\end{figure}

The partial information decomposition framework describes two distinct ways that different identities can be ``entangled:" they can both individually communicate the same information (redundancy) or they can jointly communicate information that is not not disclosed by any simpler combination of sources. Following \cite{gutknecht_bits_2020} and \cite{makkeh_introducing_2021}, we can relate the redundant and synergistic information to the logical disjunction (OR) and logical conjunction (AND) respectively. Consider the redundant information between two variables: it is information that is disclosed by \textit{either} variable. An observer could choose one of the two variables at random, observe it and only it, and learn the same information as if they had chosen the other variable. For example, when considering the information redundantly present in both race and sex about income, we ask: how would we revise our estimate of a given person's income if we knew that they were Black \textit{or} a woman? Across all 1000 resampled distributions for the 2020 dataset, the overall expected income was \$43,304.67, however if we restrict our analysis to only those individuals who are Black \textit{or} a woman (excluding White men but including Black men, White women, and Black women), we find that the expected income drops to \$39,006.65: this difference of \$-4,298.01 is the expected income penalty associated with having \textit{at least} one of the identities in question, without knowing explicitly which one the person had. To consider the intersectional synergy, we can also ask what the expected income for someone who is both Black \textit{and} a woman is: \$35,908.84. This is a \$7,395.81 financial penalty specific to black women and another \$3097.80 penalty that goes above and beyond being either Black or a woman. If we consider the Venn Diagram in Figure \ref{fig:pid_venn}, the shared cost of being Black or a woman is represented by the innermost, hatched, intersection, while the specific cost of being Black and a woman is represented by the outermost oval of synergy. 

Formally:
\begin{eqnarray}
    \textbf{E}[Income] &=& \$43,304.67 \notag \\
    \textbf{E}[Income | Black \vee Woman] &=& \$39,006.65 \notag \\
    \textbf{E}[Income | Black \wedge Woman] &=& \$35,908.84 \notag 
\end{eqnarray}

You can also do the same analysis considering privileged identities such as Whiteness, masculinity or the intersection of whiteness and masculinity. Consider the logical disjunction of Whiteness and masculinity:

\begin{eqnarray}
    \textbf{E}[Income | White \vee Man] &=& \$45,775.02 \notag \\
    \textbf{E}[Income | White \wedge Man] &=& \$56,212.66 \notag 
\end{eqnarray}

Being White \textit{or} male comes with an additional benefit of \$2,470.34, and the added benefit of being White and male above and beyond even that is \$10,437.64. While being Black and woman comes with an intersectional ``cost" of $\approx\$3,000$, being White and a man comes with an intersectional ``reward" of over $\$10,000$. This shows that there are multiple ways that intersecting identities can interact, which we might call ``redundant intersectionality" vs. ``synergistic intersectionality." The redundant intersectionality is the set of experiences or effects shared by both identities that is common to both of them. In contrast, the synergistic intersectionality is the extra effect specific to the identities in question that goes ``above and beyond" the shared experiences of both independent identities. For visualization of this, see Figure \ref{fig:cost-benefit}.

\subsection{Robustness of the Pipeline}
\begin{figure}[!h]
    \centering
    \includegraphics[scale=0.4]{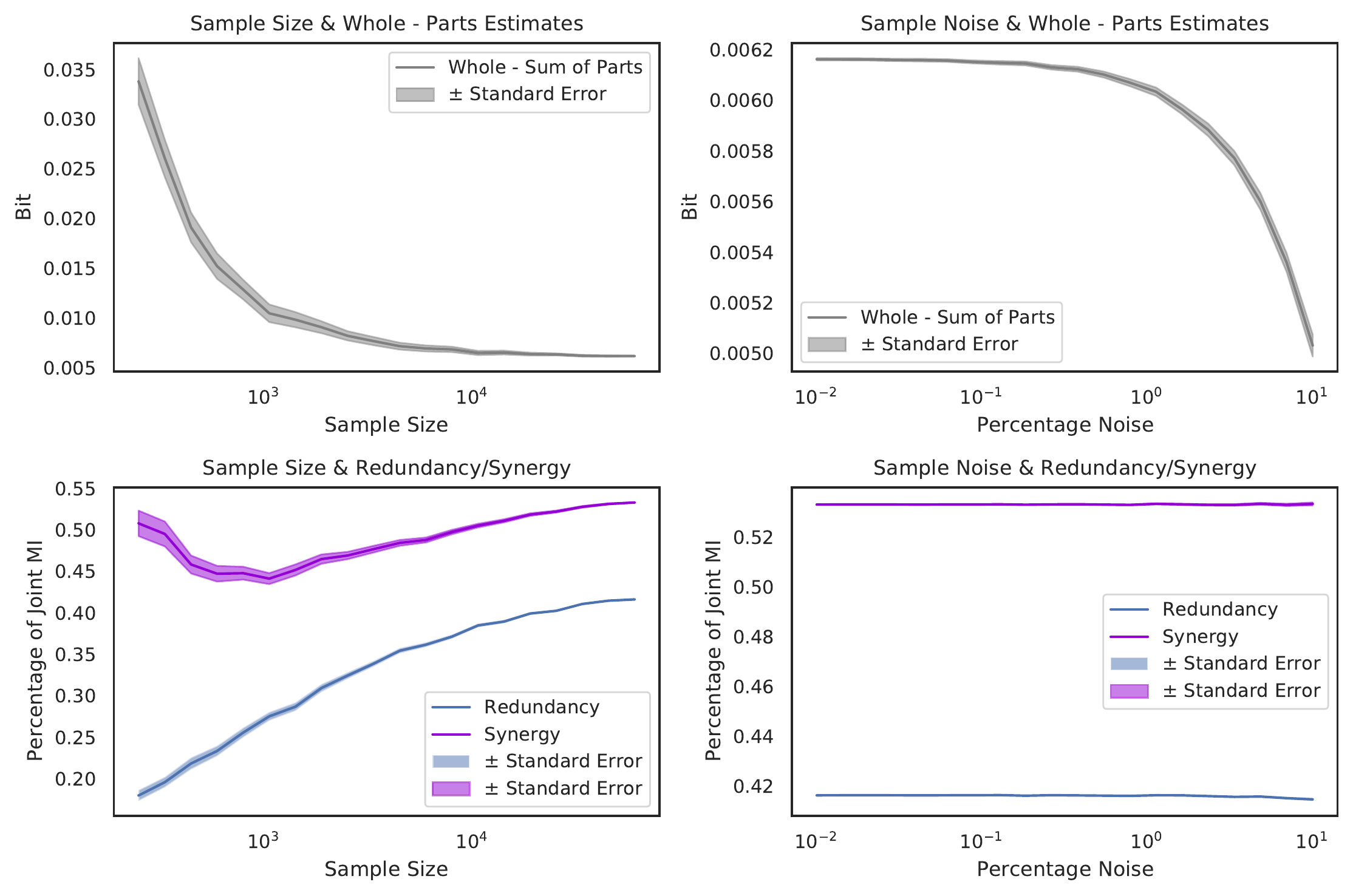}
    \caption{\textbf{Robustness of the PID pipeline to noise and small sample sizes.} The four plots show how the analytical pipeline described above performs when the initial dataset is shrunk, or has noise added to it. $I(Race, Sex ; Income)$ was used as an example data set, although comparable results can be seen for the other relationships. \textbf{Top Left} Assessing how the difference between the joint mutual information and the sum of the marginal mutual informations changes as the number of samples in the original data set is reduced. We can see that smaller samples tend to over-estimate the difference between whole - sum, although convergence happens reasonably quickly. \textbf{Top Right} Plotting the difference between the joint MI and the sum of the marginals as progressively more noise is added to the data.  We can see that the result is somewhat sensitive to noise; with 10\% noise in the data, the percentage change in whole-minus-sum information being $\approx-18.355\%$. \textbf{Bottom Left} The proportion of the total MI that is redundant or synergistic as the sample size increases. While the absolute values and relative ratios of the particular atoms change, the pipeline is always able to identify the existence of synergistic relationships in the data. \textbf{Bottom Right} The proportion of MI that is redundant or synergistic as the noise increases. Despite the change in whole-minus-sum values as noise increases, the relative ratios remain largely constant. These result show that the PID inference pipeline is reasonably robust both to decreased sample sizes and noise in the data.}
    \label{fig:size_noise}
\end{figure}

We stress-tested our pipeline to assess how robust it might be to natural limitations in data collection: smaller sample sizes and noise in the data. By virtue of working with census data, we naturally have access to a larger-than-usual data set, although it was uncertain how well this kind of information-decomposition approach would work on smaller data sets. To test this, we re-ran the whole inference for the $I(Race, Sex ; Income)$ analysis, using 20 different-sized subsets of the original data set each time. Subset sizes were arranged logarithmically, from 250 samples to the full-sized data set (59858 samples). For each subset, we randomly sampled individual respondents, and replicated each subset 600 times, to create a distribution for each subsample size. We found that, for very small subset sizes, there was a significant \textit{over-estimation} of the difference between the joint mutual information and the sub of the marginals, although the estimates declined rapidly and converged towards the true value by $\approx$ 1000 samples (see Fig. \ref{fig:size_noise} Top Left). When looking at the redundancy and synergy atoms directly (normalized by the joint mutual information to give a proportion of the total information), we found that all subsample sizes found strong evidence of synergies between race and sex on income. The specific values changed, as did the relative ratio between them, but this is a promising result that, in general, the identification of intersectional synergies is possible even in a comparatively smaller dataset (see Fig. \ref{fig:size_noise} Bottom Left).

To assess the effect of noise in the data, we took a similar approach. We re-ran our analysis, each time randomly permuting increasingly large subsets of the initial dataset. The 20 subset sizes were arranged logarithmically from 0.01\% to 10\%, and each subset size was re-tested 600 times. We found that that the whole-minus-sum heuristic showed a modest decrease as the randomization increased (see Fig. \ref{fig:size_noise} Top Right), however the percentage of the total mutual information that was redundant and synergistic remained extremely stable (see Fig. \ref{fig:size_noise} Bottom Right). These results collectively provide evidence that the PID analysis pipeline can be used for smaller, and noisier datasets while retaining sensitivity to the presence of higher-order statistical synergies. 

\subsection{Comparing PID and Linear Regression with Interaction Terms}

\begin{figure}[!h]
    \centering
    \includegraphics[scale=0.4]{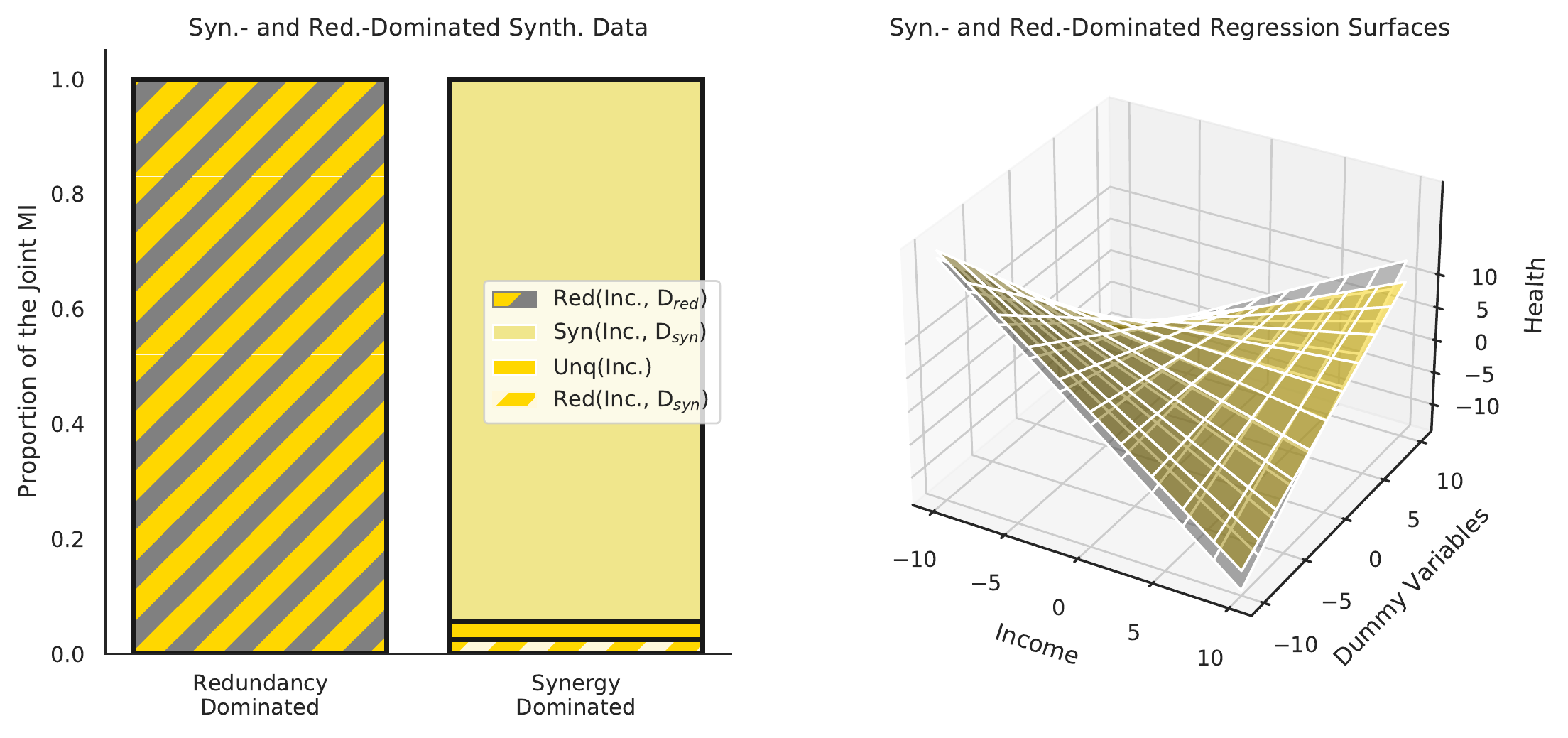}
    \caption{\textbf{Linear regression with interaction terms fails to differentiate between synergy- and redundancy-dominated relationships.} \textbf{Right:} The partial information decomposition for two different dummy datasets. $I(Income, D_{red} ; Health$ was constructed to be completely redundant: Income and $D_{red}$ disclose exactly the same information about Health. In contrast $I(Income, D_{syn} ; Health$ was constructed to be largely synergy dominated. Despite these clear differences in structure, the linear regressions with multiplicative interaction terms return essentially the same regression (see the surface plots on the right). This shows that linear regressions cannot uncover synergistic higher-order intersections of the sort that the intersectionality framework focuses on.}
    \label{fig:dummies}
\end{figure}

As discussed in the Introduction, the current gold-standard methodology for assessing intersectionality in quantitative data is the use of multiplicative interaction terms in linear regression \cite{rouhani_intersectionality-informed_2014}, however the linear interaction term fails to effectively disambiguate between different kinds of relationships between variables. To demonstrate this, we create two dummy datasets based on the empirical data (taken from the 2020 ASEC data), one of which is completely redundant, and another which is almost completely synergy-dominated, and show that the linear regressions (including interaction variables) are practically the same, despite the enforced differences in interaction structure. We argue that this indicates that regression-based approaches to intersectional data analysis have been missing important structures and relationships in the data that have clear implications for intersectional analysis.  

We began with the relationship between Income and Health, known to be correlated in the real data. To create a triad dominated by redundancy, we created a dummy variable $D_{red}$ constructed in such a way that the information it contained about Health Status was completely redundant with Income:

\begin{equation}
    D_{Red} = (1+Income) \textnormal{ mod } 4
\end{equation}

$D_{red}$ is functionally a copy of the Income dataset, with every value incremented by 1 (and wrapping around for values higher then the maximum income. Our joint mutual information decomposition then becomes:

\begin{equation}
    I(Income, D_{Red} ; Health)
\end{equation}

When we do the full partial information decomposition using the Williams and Beer redundancy function $I_{min}$, we find that, as expected, all of the joint mutual information is redundant (the results are functionally the same when using $I_{BROJA}$, see Supplementary Material.

The second dummy variable was constructed to result in a synergy-dominated decomposition and is given by:

\begin{equation}
    D_{Syn} = (Income + Health) \textnormal{ mod } 2
\end{equation}

The operation is analogous to a generalized logical XOR operator (see \cite{rosas_reconciling_2020} for reference). We decompose:

\begin{equation}
    I(Income, D_{syn} ; Health)
\end{equation}

to produce a partial information decomposition that is 94.4\% synergistic, with the remaining 5.6\% of the information distributed over unique and redundant partial information atoms (it is impossible to create a purely synergistic relationship involving Income and Health Status, since Income already discloses information about Health Status and they are therefore not independent). 

When we feed these two datasets into standard linear regressions with multiplicative interaction terms, we get:

\begin{eqnarray}
    \hat{Health} &=& 2.9 - \notag \\ 
    &&(0.34\times Income) - \notag \\ 
    &&(0.34\times D_{red}) + \notag \\
    &&(0.13\times Income\times D_{red}) \notag
    \label{eq:reg_red}
\end{eqnarray}

and 

\begin{eqnarray}
    \hat{Health} &=& 2.6 - \notag \\ 
    &&(0.29\times Income) - \notag \\
    &&(0.17\times D_{syn}) + \notag \\
    &&(0.15\times Income\times D_{syn}) \notag
    \label{eq:reg_syn}
\end{eqnarray}

Looking at these regressions, it is clear that, despite the completely different information-structures, the linear regressions with multiplicative interaction terms finds very similar statistically significant coefficients. Even a more involved analysis such as a Shapely decomposition would be unable to distinguish the ``types" of interaction, instead treating it as a lump sum. These results can be visualized in Figure \ref{fig:dummies}. This is a significant finding because it shows that linear regression with multiplicative interactions terms is incapable of distinguishing between ``true/synergistic" intersectional relationships where the joint effects of identities are greater than the sum of their parts, and ``redundant" interactions between identities that are not non-additive in the way originally discussed by Crenshaw and Bailey. 

\section{Discussion}

In this work, we have shown how the formal frameworks of information theory writ large, and partial information decomposition specifically can be used to address the question of intersectionality in data. Using a decade of data from the US Census Annual Social and Economic Supplement, we show that synergistic, ``greater-than-the-sum-of-their-parts" interactions can be identified in the interactions between identities such as race, sex, and class on outcomes such as income and general health status and that the relative proportions of redundant, unique, and synergistic information remain stable across the decade. We furthermore show that established ``gold-standard" analyses such as linear regression with interaction terms fails to discriminate between synergistic and redundant modes of information sharing. This strongly suggests that salient social dynamics and relationships are being missed by many current approaches and that complementary frameworks may be illuminating. This work fits into the broader project first outlined by Abbott in the seminal critique \textit{Transcending General Linear Reality} \cite{abbott_transcending_1988}. Abbott argues that over-reliance on general linear models has influenced how researchers think about the world: that the map (consisting largely of linear relationships between interacting entities, assumptions about single effects being generated by single causes, etc.) is confused for the territory (the real world, which is highly non-linear and admits complex, potentially higher-order causal interactions). The results presented here show that the standard practices that make the assumptions Abbott describes are missing potentially important relationships in empirical data: the existence of intersectional synergies are direct evidence of higher-order interactions between attributes that are both irreducible and, by virtue of the use of the effective information, suggestive of a causal relationship. 

Beyond technical and methodological advances, we feel that the conceptual distinction between ``redundant" and ``synergistic" interactions may be of theoretical interest. For example, some authors have suggested that intersectionality as a framework balkanizes individuals into small groups with limited shared solidarity. This concern is exemplified by by Naomi Zack, who writes:

\begin{quote}
    ``...as a theory of women's identity, intersectionality is not inclusive insofar as members of specific intersections of race and class can create only their own feminisms." \cite{zack_inclusive_2005}
\end{quote}

Zack goes on to argue:

\begin{quote}
    ``These ongoing segregations make it impossible for women to unite politically and they have not ended exclusion and discrimination among women, especially in the academy." \cite{zack_inclusive_2005}
\end{quote}

Without weighing into political dimension of Zack's argument, we claim that the partial information decomposition shows that ``exclusive" intersections and ``shared oppression" can co-exist in society. Continuing the example of misogynoire as a relevant case, the partial information results suggests that a given individual who is both Black and a woman will simultaneously experience: synergistic costs associated with pure misogynoire (corresponding to the synergistic partial information atom), ``generic" anti-black racism and misogyny (corresponding to the unique partial information atoms), as well as the ``shared" costs experienced by all those who are either Black or a woman (including Black men and White women). The same analysis can be done with regards to the rewards of privilege. An interesting question for further research would be how privileged and marginalized identities interact: for conceptual simplicity we have focused on the two cases where identities ``stack" in the same direction (Blackness and womanhood are both generally marginalized, while Whiteness and masculinity are generally privileged), although the synergies that may emergence when privileged and marginalized types of identities co-exist coult yield interesting novel insights. 

We should stress that the framework presented here is not intended to either ``prove", or ``disprove", the existence of intersectionality as a theory. Instead, it is a statistical framework that can be used to identify irreducible intersectional synergies in large data sets. We hope that this may inform evidence-based discussion around social issues. The failure to find a particular synergistic relationship in a particular data set should \textit{not} necessarily be grounds to claim that a particular intersection is ``unreal" or ``unimportant" given the restrictions inherent in working with limited data sets and limitations of data collection. By the same token, the identification of an unexpected statistical synergy should be addressed critically and assessed in the context of previous scholarship. 

Despite its utility, information theory comes with some particular limitations that must be addressed. The first is that the amount of data required for a reliable inference is much larger than what is required for a linear regression. This is particular pressing for higher-order analyses that require reliable estimation of potentially high-dimensional joint-probability spaces. For studies that rely on ``big data" sets (e.g. Census data, data gathered from social media, etc), this is unlikely to present a problem, although smaller-scale survey-based studies of local populations may not generate sufficient data, in which case linear methods may be preferable. Our stress-testing of our pipeline can allay those concerns somewhat, although there are known lower-bounds on the ability to reliably infer probability distributions from finite datasets \cite{bossomaier_introduction_2016}. Another pressing issues is that the number of partial information atoms grows super-exponentially with the number of sources: for example, assessing the intersections of 5 distinct identities (e.g. race, sex/gender, class, sexuality, ability) on a single outcome (e.g. life-expectancy) would require computing 7,828,352 distinct values (many of which would be very hard to interpret higher-order terms such as the information about life expectancy disclosed by the joint state of race and sex/gender or class or the joint state of sexuality and ability). Given this difficulty, other frameworks for information decomposition may be worth exploring, such as the ``synergy-first" proposal given by Quax et al., \cite{quax_quantifying_2017} or Rosas et al \cite{rosas_operational_2020}. 

The relative simplicity of applying the partial information decomposition to the effective information opens the doors to a range of future research directions. For instance, it may be worthwhile to revisit previous empirical studies of intersectionality (e.g. those discussed in \cite{rouhani_intersectionality-informed_2014} to disambiguate which interactions are synergistic in nature, vs. which ones are redundant in nature. It also suggests that researchers doing a multiplicative-interaction based analysis of intersectional relationships may consider supplementing their analysis with a PID-based framework to explicitly untangle redundant and synergistic components. Alternately, being able to untangle what features are redundantly predictive of an outcome, vs. which are synergistic can provide deeper insights into the generative dynamics of social relationships. Essentially any analysis that has previously relied on multiple linear regressions for mediator analysis is essentially fair-game, and there may be a considerable number of hitherto-unrecognized higher-order interactions to be explored and understood. 

\section{Conclusions}
Despite the extensive and valuable work that has been done integrating intersectional frameworks in data analysis and social science research, the most commonly-used methods (e.g. linear regression with multiplicative interaction terms) fail to capture the range of ways that different identities can interact and predict outcomes. Information theory provides an appealing alternative framework that allows users to identify different modes of interactions of intersecting identifies, while simultaneously shedding the requirements imposed by assumptions of linearity. These modes, which we called ``redundant intersections" and ``synergistic intersections" correspond to shared costs/benefits of multiple interacting identifies and the exclusive costs/benefits respectively. Examples can be seen in the redundant and synergistic interactions between race and sex considered jointly, allowing the identification of a strong synergistic cost of misogynoire, as well as a shared common cost between Blackness and womanhood. We hope that the framework detailed here will enable novel and insightful work on intersectionality in empirical data.   

\section*{Author Contributions}
TFV and PK conceptualized the project. TFV performed data analysis, and drafted the manuscript. PK assisted with interpretation and editing the manuscript.

\acknow{TFV and PK are supported by the NSF-NRT grant 1735095, Interdisciplinary Training in Complex Networks and Systems at Indiana University Bloomington. We would like to thank Dr. Byungkyu Lee and Dr. Michael Schultz for their detailed feedback and discussion of the early versions of this manuscript. We would also like to thank Zackary Dunivin, and Bradi Heaberlin for thoughtful discussions about intersectionality, statistics, and the concept of this project. Finally, we would like to thank Dr. Olaf Sporns for advice and support. }

\showacknow{} 

\bibliography{pid_intersectionality}

\end{document}


\maketitle

\begin{figure}[!h]
    \centering
    \includegraphics[scale=0.5]{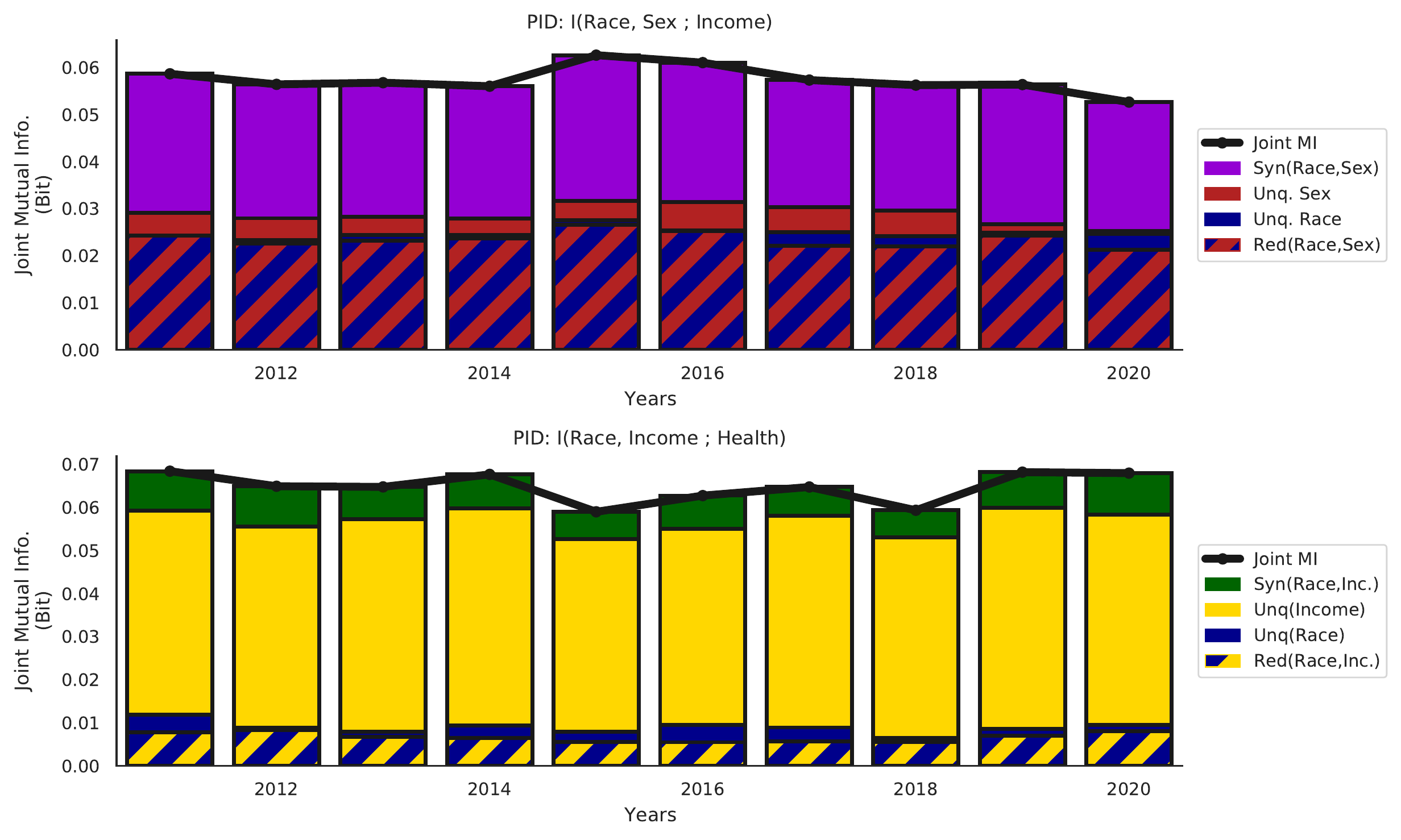}
    \caption{\textbf{The PIDs for \textit{I(Race, Sex ; Income)} and \textit{I(Race, Income ; Health)} using the $I_{BROJA}$ function.} We can see that, while the values are not exactly the same, the two different functions for calculating the PIDs ($I_{min}$ from \cite{williams_nonnegative_2010} and $I_{BROJA}$ from \cite{bertschinger_quantifying_2014,griffith_quantifying_2014}) find roughly the same distributions of redundant, unique, and synergistic information. This suggests that our analysis is robust to free parameters.}
    \label{si_fig:broja_bars}
\end{figure}

\begin{figure}[!h]
    \centering
    \includegraphics[scale=0.5]{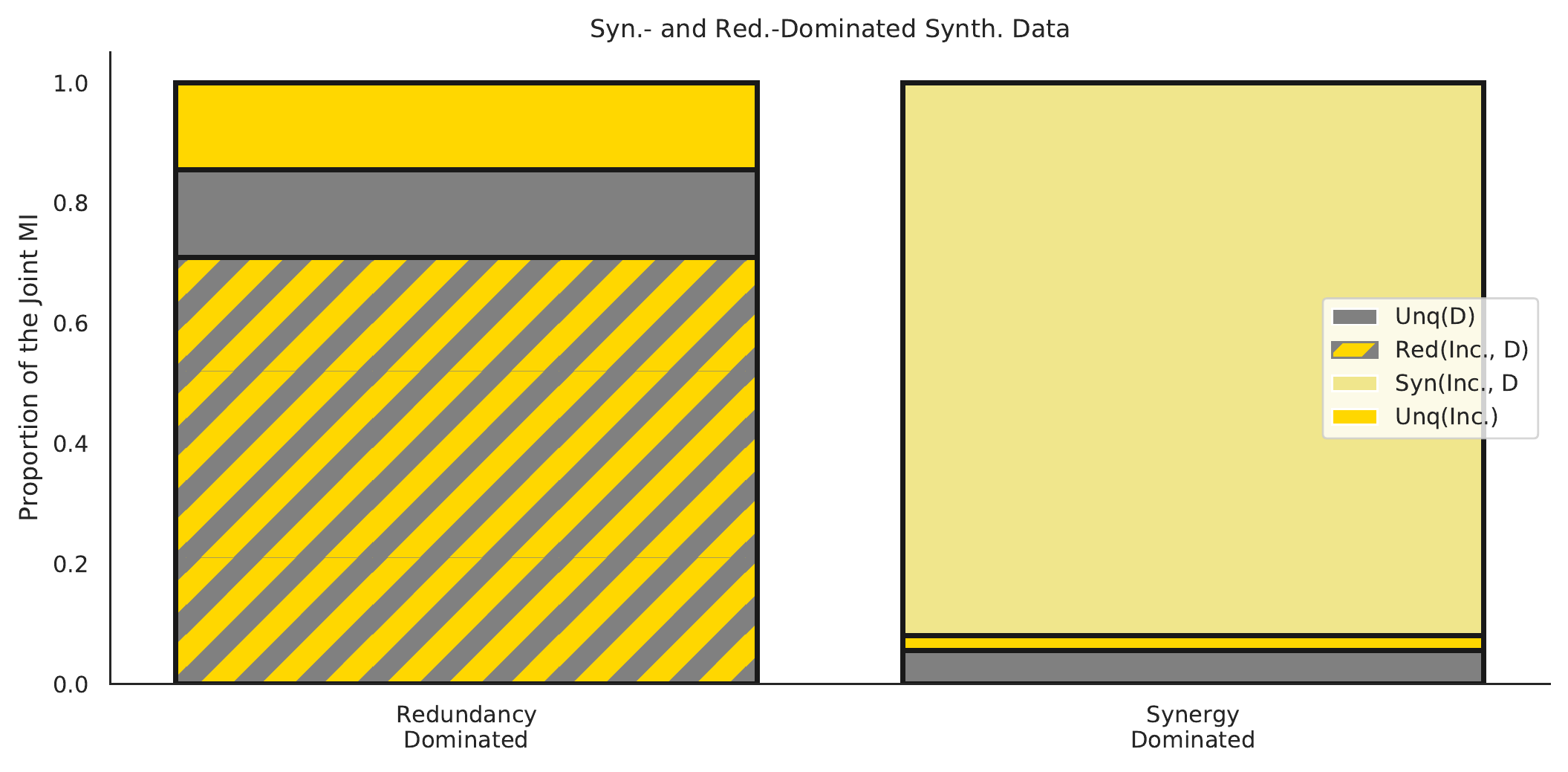}
    \caption{\textbf{The PIDs for the synergy and redundancy dominated synthetic datasets using the $I_{BROJA}$ function.} As with the empirical data, the results of the dummy decomposition are functionally the same regardless of whether one is using $I_{min}$ or $I_{BROJA}$.}
    \label{si_fig:broja_dummies}
\end{figure}

\bibliography{pid_intersectionality}